\documentclass[sigconf]{acmart}

\renewcommand\footnotetextcopyrightpermission[1]{} 
\usepackage{balance}
\usepackage{multirow}
\usepackage{makecell}

\AtBeginDocument{
  \providecommand\BibTeX{{
    \normalfont B\kern-0.5em{\scshape i\kern-0.25em b}\kern-0.8em\TeX}}}

\copyrightyear{2022}
\acmYear{2022}
\setcopyright{acmcopyright}
\acmConference[]{July}{2022}{ChoreoGraph}

\acmSubmissionID{254}

\begin{document}

\title[ChoreoGraph]{ChoreoGraph: Music-conditioned Automatic Dance \\Choreography over a Style and Tempo Consistent Dynamic Graph}

\author{Ho Yin Au$^1$, Jie Chen$^{1,*}$, Junkun Jiang$^1$, Yike Guo$^{1,2}$}
\affiliation{
\institution{$^1$Department of Computer Science, Hong Kong Baptist University, Hong Kong SAR, China}
\institution{$^2$Data Science Institute, Imperial college London, London, UK}
\institution{\{cshyau, csjkjiang, chenjie, yikeguo\}@comp.hkbu.edu.hk,$~$yg@doc.ic.ac.uk}}

\thanks{$^*$ Corresponding author: Jie Chen}

\renewcommand{\shortauthors}{Au et al.}

\begin{abstract}
To generate dance that temporally and aesthetically matches
the music is a challenging problem, as the following factors need to be considered. First, the aesthetic styles and messages conveyed by the motion and music should be consistent. Second, the beats of the generated motion should be locally aligned to the musical features. And finally, basic choreomusical rules should be observed, and the motion generated should be diverse. To address these challenges, we propose ChoreoGraph, which choreographs high-quality dance motion for a given piece of music over a Dynamic Graph. A data-driven learning strategy is proposed to evaluate the aesthetic style and rhythmic connections between music and motion in a progressively learned cross-modality embedding space. The motion sequences will be beats-aligned based on the music segments and then incorporated as nodes of a Dynamic Motion Graph. Compatibility factors such as the style and tempo consistency, motion context connection, action completeness, and transition smoothness are comprehensively evaluated to determine the node transition in the graph. We demonstrate that our repertoire-based framework can generate motions with aesthetic consistency and robustly extensible in diversity. Both quantitative and qualitative experiment results show that our proposed model outperforms other baseline models.
\end{abstract}


\maketitle
\section{Introduction}
\vspace{1mm}
Human dance is a performing art that is composed of human pose and motion sequences by choreographers based on background music or the theme of the performance. This human activity encodes human cultures, expressions, and beat information, which should be aligned with the paired songs or music. The dance style of the dance sequences can be informative, encoding meaningful expressions to represent the theme of the performance and perform rituals like Classical Ballet for Swan Lake and Ceremonial Dances, or can be groovy and unremitting to represent dancers' skills and emotions like Street Dances and Contemporary Dances. To design a dance sequence that is appropriate for the background theme and music, the dance choreographer requires professional experience and extensive reasoning ability to handle and encode the complex relationship between dance styles, poses meanings, and music rhythms into the dance choreography.

Recently, the availability of stable and accurate 3D motion capture systems provides opportunities for dance performers and technological companies to record enormous amounts of paired dance and music data, dance generation based on data-driven approaches becomes the hot trend of research \cite{li2021learn,valleperez2021transflower,wen2021autoregressive,li2017autornn}. These approaches can capture the correlation between dance styles, poses meanings, and rhythms statistically based on the recorded music and dance data, and apply the learned relationship to generate new dance sequences that match the input variables such as music and dance styles. Although the quality of generated dance sequences from the trained generative models is still not decent from human evaluation, those models still help multimedia content creators and dance choreographers to design new dance moves for non-uniform music.

On the other hand, traditional methods \cite{bowden2000tradition, min2012motiongraph,kovar2008motiongraph,heck2007motiongraph} that generate motion sequences based on re-aligning sliced motion segments also benefit from the increased motion and music data. More representative and informative feature extraction methods are implemented based on the data statistics, and more distinct and stylish motions are inserted into the collection for those methods to choose in the choreograph process. Those methods can produce realistic style-matched motion from the extracted features guided by professional artists, but they also suffer from generating motion with matched rhythmic features because motion segments with perfect style and rhythm are not always available in the collection. Also, the problem of action completeness exists in the segment selection methods, which cause unpleasant artifacts such as sudden stop in spinning or waving motion in the generated motion sequence.

While the state-of-the-art generative models perform well on generating style consistent motion with smooth transitions \cite{valleperez2021transflower}, they usually struggle on music with immense style change. The generated motion do not response to such style change, showing limited style variation and producing repetitive moves. Although, traditional methods can handle music with immense style change that is captured by designed features, the rhythm of the generated motion usually cannot match that of the music.
We want to design a framework that is based on the style and rhythm information extracted using data-driven approaches, while preserving the ability to generate realistic motion with immense music style change in traditional methods. Therefore, we design a framework that contains Style Embedding modules and Tempo-Density estimators trained using data-driven approaches, while having a Motion Dynamic Graph for motion node selection and final motion generation.

In this work, based on the input motion segment database and input music, we propose a motion generation framework that contains Style Embedding modules that are responsive to music style change, Dynamic Time-Warper which can re-align the motion segment according to music rhythmic features extracted from Tempo-Density module, and Dynamic Graph which construct final motion sequence by choosing warped motion segments after taking music style, action completeness, and motion transition into consideration. The Style Embedding modules and Tempo-Density estimators are trained using data-driven approaches, while the Dynamic Graph design is based on the traditional method Motion Graph \cite{kovar2008motiongraph,safonova2007motiongraph,min2012motiongraph,heck2007motiongraph}.

To summarize, our contributions will be as follows:
\begin{enumerate}
    \item   We propose a motion generation framework which select and warp motion segments based on trained Style Embedding and Tempo-Density modules, while synthesize output motion that is responsive to music style change and action completeness.
    \item   We propose a data-driven learning strategy to extract motion rhythmic feature based on motion distortion. The extracted motion Tempo-Density allows efficient motion segment warping for music rhythm and tempo alignment.
    \item Based on the Dynamic Graph, we have incorporated choreomusical considerations such as action completeness, transition naturalness, and style appropriateness into the choreography pipeline. User study shows dance generated by our ChoreoGraph is more natural and diverse.
\end{enumerate}

\vspace{2mm}
\section{Related Work}
\vspace{1mm}
\subsection{Deep Learning Method for Motion Synthesis}
Recent works on motion synthesis mainly focus on deep learning methods such as CNNs \cite{holden2016deep}, RNNs \cite{li2017autornn, fragkiadaki2015rnn,jain2016rnn,du2019rnn,chiu2019rnn}, Normalizing Flow Models \cite{valleperez2021transflower}, and Transformers \cite{li2021learn,aksan2020transformer,bhattacharya2021transformer2,petrovich2021transformer,li2021dancenet3d, li2020learning}, which can correlate cross-modality features and extract meaningful representations from complex and diverse real-world data and produce motion sequences better satisfying the complex cross-modality alignment requirement. Especially, music conditioned dance generation task requires the model to learn, extract, and utilize cross-modal information, which is very abstract and ambiguous.

AI Choreographer \cite{li2021learn} and Transflower \cite{valleperez2021transflower} propose end-to-end data-driven learning approaches, extracting cross-modal information from elemental music features and human skeleton motion data to generate output motion sequences using Transformer \cite{vaswani2017transformer, bhattacharya2021transformer2} or Normalizing Flow \cite{rezende2015normalizingflow} model. These models can generate motion sequences satisfying multiple constraints, especially on audio beat alignment and style matching. However, those methods usually overfit the training data, and the generated dance contains unnatural and unattractive patterns. When the audio beat and rhythm go very fast, the generated motions become monotonous and repetitive as the model wants to match every audio beat.

From our observations on dance sequences designed by professional choreographers, not all audio beats in the music should be matched with motion beats, as the number of motion beats is usually smaller. Instead, every motion beat in the dance should be matched with audio beats. The misaligned motion beats will create awkward dissonance from the human perspective.

In our work, we select and warp sequences from the motion dataset to make sure every motion beat has a matched audio beat. This guarantees both the naturalness and beat alignment of the output motion. As the selection and warping are based on data-driven learning approaches, the output motion is aesthetically aligned to conditioned music while matching the music-motion style.

\subsection{Time Series Alignment}
Motion re-tempo and beat alignment can be viewed as the one-dimensional time series alignment problem if the motion and music rhythms can be expressed as one-dimensional sequences. Dynamic Time Warping \cite{sakoe1978dynamic} is one of the most representative and powerful methods for aligning two time series and evaluating the alignment cost. Soft DTW \cite{cuturi2017soft} and DILATE \cite{le2019shape} turn DTW from a non-differentiable dynamic programming loss into a differentiable alignment loss, which provides alignment evaluations between the prediction and ground truth time series in neural network training. The temporal loss in DILATE \cite{le2019shape} provides extra focus on temporal change detection, which is ideal for the music-motion beat alignment tasks. However, such differentiable alignment loss prevents explicit alignment path evaluation and contains boundary alignment assumption, which prevents its application to subsequence extraction.

In our model, the Dynamic Time-Warper module is implemented based on Dynamic Time Warping \cite{sakoe1978dynamic} subsequence extraction in Librosa library \cite{mcfee2015librosa}, which takes a shorter music tempo-density time series and a longer motion tempo-density time series, and generates an alignment path for subsequence extraction from motion.

\begin{figure*}[t]
\centering
\includegraphics[width = 1\linewidth]{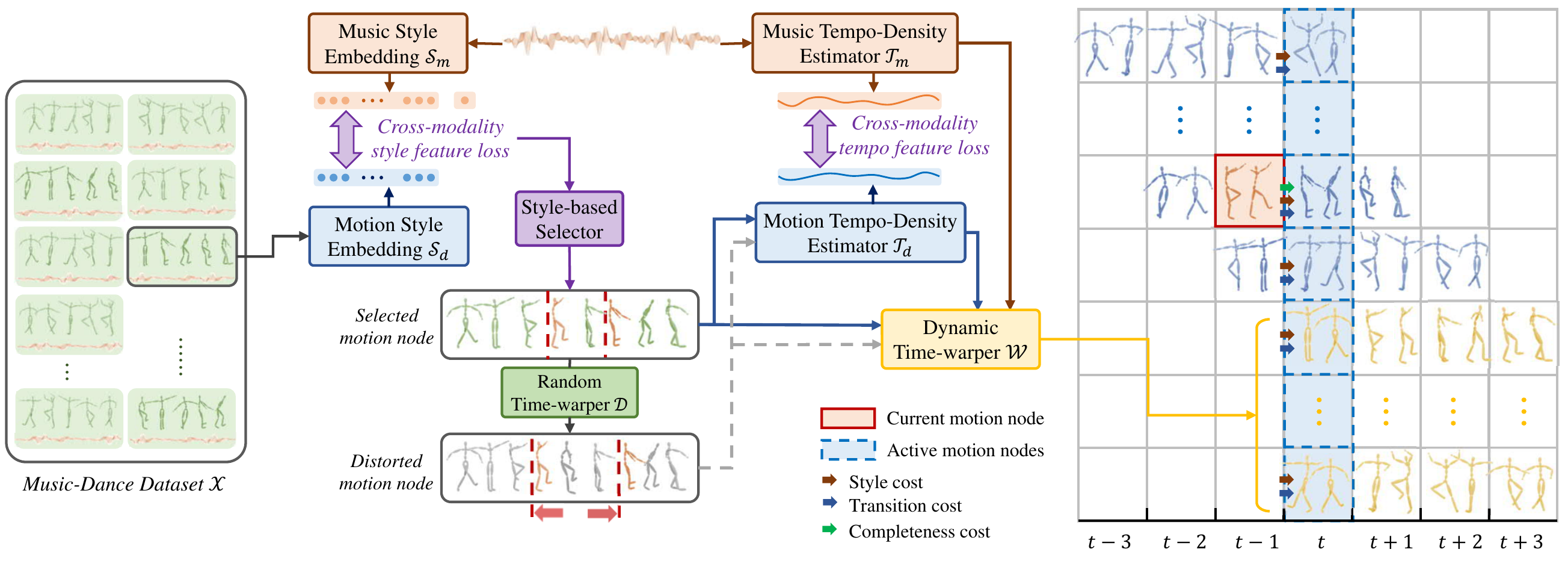}
\caption{Proposed system diagram of our motion generation model.}
\label{fig:system}
\end{figure*}

\subsection{Graph-based Motion Synthesis}
Instead of generating motion poses from the trained models, dance sequences can also be generated using traditional methods like Motion Graph \cite{min2012motiongraph,kovar2008motiongraph,safonova2007motiongraph}, statistic feature based models \cite{bowden2000tradition,pullen2000tradition, shiratori2006tradition}, and heuristic-based algorithm approaches \cite{tendulkar2020heuristic} by searching for appropriate motion segments in the dataset. Yang et al. \cite{yang2020statistics} applied a graph-based framework based on the stochastic greedy search to synthesize body motions for social conversations. ChoreoMaster \cite{chen2021choreomaster} combines the data-driven methods and the traditional Motion Graph approach by training a choreomusical embedding network to extract and process cross-modal information into style and rhythm requirements, and examining those requirements in the motion node selection phase in the graph-based framework.

When combining music information with these traditional methods, the custom-designed cross-modality cost will usually be added to the framework. These methods usually exploit some simple and intuitive relationship between music and motion such as rhythm alignment, but they usually fail to capture abstract correlations such as style and theme accurately. Also, these methods need to scan through an enormous amount of motion node candidates to find a motion segment satisfying all the conditions. Therefore, these methods either need special labeling in the motion dataset for effective searching or have a coarse resolution of the rhythm feature to avoid the exponential growth of graph connections. For example, ChoreoMaster \cite{chen2021choreomaster} categorizes motion rhythm into 13 common 2-second rhythm signatures, and converts beat alignment constraints into a classification problem of 13 classes. The categorization of motion rhythm imposes a strong limitation on the beat pattern in the generated motion and prevents the application of the framework to non-uniform music.

In our work, only a small dataset is needed for selection as the music rhythm alignment requirement will be satisfied by motion segment warping. As we warp the motion from the Style-based Selector to align with the music beat and rhythm, only a small number of style-matched and beat-aligned motion nodes will be inserted into the graph, and this makes the search problem easier. Also, those nodes will only be activated once, and the outdated nodes will be removed from the graph to reduce computation costs.

\section{Proposed Method}
\vspace{1mm}
Our method is composed of the learning-based Tempo-Density module and Style Embedding module for motion selection and warping to produce motion candidate nodes, and a Dynamic Graph to select the best node while taking music and the last selected motion node into consideration. Our Motion/Music Style Embedding module and Tempo-Density prediction module are both trained on music-dance data pairs which samples and then time-warps motion sequences into graph nodes based on extracted music style and tempo features. After that, from all the tempo-adjusted and beat-aligned motion node candidates in the graph, our Dynamic Graph will select the best motion node by considering and programming re-examining music-motion styles and choreomusical rules over the graph edge cost computation. Given a $T$ seconds music piece, our goal is to generate a dance sequence that has the same time duration and matches the music local features and aesthetic styles.

We adopt a Dynamic Motion Graph framework as shown in Fig. \ref{fig:system}, which choreographs new dances based on source motion segments organized as 8-second clips in a database $\mathcal{X}=\{X_i\}_i$. Note that each motion clip is a choreographed dance piece according to a music clip, and the two form a data pair $(X_i, M_i)$. $\mathcal{X}$ can be created from external sources such as the AIST++ dataset \cite{li2021learn}.

We represent dance as a sequence of poses (sampled at 20 frames per second) and the music as a sequence of \textit{spectral features} (sampled at 60 frames per second). We process both the music and motion sequences based on 1-second units which we refer to as music/motion \textit{nodes}. Each node at a given time $t$ belongs to a 4-second \textit{segment window} based on which local features are extracted for style matching.
For each 4-second music segment, a Music Style Embedding Module $\mathcal{S}_m$ is designed to encode the music features into a hidden style vector $s_m$, which will be compared against the motion style vectors $\{s_d\}$ extracted for the 8-second motion clips in the Music-Dance Dataset $\mathcal{X}$ by the Motion Style Embedding Module $\mathcal{S}_d$.
Then, a Style-based Selector picks $K$ motion clips from $\mathcal{X}$.
Next, the Music and Motion Tempo-Density Estimators, i.e., $\mathcal{T}_m$ and $\mathcal{T}_d$, estimate both the music and motion tempo-densities based on which the Dynamic Time-Warper $\mathcal{W}$ works to extract the beats-aligned and tempo-adjusted 4-second motion segment $\lceil X\rfloor$ from the selected 8-second clips $X$ and organize as four new motion nodes $\lceil X\rfloor= \{x_j\}_{j=1}^4$ (1 second each) into the Dynamic Graph.
The Dynamic Graph will organize all the inserted motion nodes and activate those at the appropriate time.
For every second, one motion node $\mathbf{x_t}$ from all the activated nodes will be selected based on the graph edge cost which includes style cost, action completeness cost, and motion transition cost. Finally, the sequence of selected motion nodes will be concatenated and blended into output motion $\hat{X}$.

In the following, Sec. \ref{sec:selector} introduces the Style-based Selector for the source motion segment selection from $\mathcal{X}$. Sec. \ref{sec:embedding} introduces the Tempo-Density module for the alignment of motion to music beats and tempo. Sec. \ref{sec:dynamicgraph} describes the Dynamic Graph which organizes the time-warped motion segments into nodes and synthesizes the result motion by node transition.

\subsection{Style-based Motion Segment Selector}\label{sec:selector}

Music and dance are two highly correlated data modalities that have explicit beat alignment and implicit style connection. The style connection is often hidden in the choreographer's feelings and cultural background, which are hard to model explicitly. Therefore, we construct Motion and Music Style Embedding modules for style extraction and train them based on data-driven approaches.

For the Music Style Embedding module $\mathcal{S}_m$ and the Motion Style Embedding module $\mathcal{S}_d$ , we follow the choreomusical style embedding module design in ChoreoMaster \cite{chen2021choreomaster}. The Music Style Embedding module has 4 convolutional block layers, 2 GRU layers, and 1 linear layer. On the other hand, the Motion Style Embedding module has 4 motion graph convolution block layers \cite{yan2018stgcn}, 2 GRU layers, and 1 linear layer.
The input of the Music Style Embedding module $\mathcal{S}_m$ is a 4-second Mel-spectrogram of shape [1, 96, 4*60], and will be processed into 32-dimensional style vector $s_m$. The input of the Motion Style Embedding module $\mathcal{S}_d$ is 4-second joint coordinates of shape [3, 21, 4*20], and will be processed into a 32-dimensional style vector $s_d$. Furthermore, to select motion segments that are easier for temporal alignment in Sec. \ref{sec:embedding}, the Embedding module also has a motion beat number prediction module $\mathcal{B}_m$ composed of 4 linear layers to predict the number of motion beats in the 4-second motion sequence given $s_m$.

The Embedding modules are trained together with a paired music and motion dataset. The music mel-feature segment of 4 seconds $m \in \mathbb{R}^{4*60 \times 96}$  and motion joint coordinates segment of 4 seconds $X \in \mathbb{R}^{4*20 \times 63}$  are both fed into $\mathcal{S}_m$ and $\mathcal{S}_d$ respectively to produce music style vector $s_m = \mathcal{S}_m(m)$ and motion style vector $s_d = \mathcal{S}_d(x)$.

We consider the progression of motion and musical beats to constitute an important factor of style. We use an MLP $\mathcal{B}_m$ to predict the motion beats number based on the music style hidden vector $\mathcal{B}_m(s_m)$. The actual motion beats $N_B$ are calculated as the number of local minima of average joint velocity. The local minima are estimated using find\_peaks function in SciPy library \cite{2020SciPy-NMeth}.

Given the Embedding module outputs $s_a$, $s_m$, and $\mathcal{B}_m(s_m)$, we calculate the cross-modality style feature loss $\mathcal{L}_{style}$. Given the loss term coefficients $\lambda_{cs}, \lambda_{b} \in \mathbb{R}$, the style loss is shown below:
\begin{equation}
    \mathcal{L}_{style}(x, m) = \lambda_{cs} ||s_m - s_d||_2 + \lambda_{b} ||\mathcal{B}_m(s_m) - N_B||_2
\end{equation}

At style-based selection time, we also process the input music by the 4-second sliding window to obtain the set of music segments $\{m^t\}$, and evaluate style vectors $\{s_m^t\}$ using $\mathcal{S}_m$. On the other hand, for each 8-second motion source segment in the Music-Dance Dataset $\mathcal{X}$, we extract the middle 4 seconds to calculate style vector $s_d$ using $\mathcal{S}_d$. After that, for each $s_m^t$ in the music style vector set, We select $K=512$ motion segments from $\mathcal{X}$ by finding the best K segments with minimum style loss $\mathcal{L}_{style}$. The selected motion segment set $\{x^t_k\}_{k=1}^{K}$ for music segment $m^t$ will be processed and warped by our Dynamic Time-Warper $\mathcal{W}$ later in Sec. \ref{sec:embedding}.

\subsection{Temporal Re-alignment via Progressive Cross-modality Embedding Learning}\label{sec:embedding}
One of the most important aspects of matching between dance and music comes from the local alignment of beats and tempo. Due to the cross-modality signal differences, and the fact that there are different interpretations of what a good match is, it is difficult to set up a rule for alignment. Therefore, we propose a data-driven strategy to extract tempo-density from motion sequences and predict such motion tempo-density based on input music features.

The motion tempo-density is a one-dimensional feature that estimates the tempo of the motion segment. For motion segment $x \in \mathcal{X}$ of shape [3, 21, 8*20], the Motion Tempo-Density Estimator $\mathcal{T}_d$ extract the motion tempo-density of shape [1, 8*20]. The estimator $\mathcal{T}_d$ is trained using an autoencoder-like approach, with tempo-density and motion content vector as the hidden vectors for motion reconstruction. First, our encoder $\mathcal{T}_d$ is composed of 1 transformer layer, 5 convolution blocks, 5 deconvolution blocks, and 2 linear layers. The input motion will first be processed into a large hidden vector of size [4096] by the transformer layer and the convolution blocks. Then, the 5 deconvolution blocks will predict the tempo-density from the hidden vector. Also, the 2 linear layers will estimate a smaller motion content vector of size [1024] from the 4096-dimensional hidden vector. The decoder $\mathcal{T'}_d$ shares the same structure as $\mathcal{T}_d$ but in reverse order, and the convolution blocks are replaced by deconvolution blocks, and so on.

For the model training of $\mathcal{T}_d$, we designed a Random Time-Warper $\mathcal{D}$ that randomly selects some motion beats, the local minima of average joint velocity, in the motion sequence $X$ and shift them and corresponding frames slightly along the time axis. The remaining frames will be linearly interpolated based on the shifted beat location. The random-warped motion sequence $\lceil X\rfloor$ should have the same motion content as the base motion sequence but have a different tempo-density due to the random-warping process. At training time, motion $X$ and warped motion $\lceil X\rfloor$ will be processed by $\mathcal{T}_d$ to obtain content vector and tempo-density $X_{td}, X_c = \mathcal{T}_d(X), \lceil X\rfloor_{td}, \lceil X\rfloor_c = \mathcal{T}_d(\lceil X\rfloor)$. Then, we reconstruct 4 motion sequences based on different combinations of content vectors and tempo-densities to evaluate reconstruction loss. Given the loss coefficient $\lambda_{cm} \in \mathbb{R}$, the loss for training $\mathcal{T}_d$ and $\mathcal{T'}_d$ is shown below:

\begin{equation}
\begin{split}
    \mathcal{L}_{td}(X, \lceil X\rfloor) = & ||\mathcal{T}_d(X_{td}, X_c) - X||_2 + ||\mathcal{T}_d(\lceil X\rfloor_{td}, \lceil X\rfloor_c) - \lceil X\rfloor||_2\\
    & + ||\mathcal{T}_d(X_{td}, \lceil X\rfloor_c) - X||_2 + ||\mathcal{T}_d(\lceil X\rfloor_{td}, x_c) - \lceil X\rfloor||_2\\
    & + \lambda_{cm} ||X_c - \lceil X\rfloor_c||_2
\end{split}
\end{equation}

After the training of $\mathcal{T}_d$, we expect the predicted tempo-density represents the temporal features of the motion, while the motion content vector carries the remaining motion related information.

After that, we build and train Music Tempo-Density Estimator $\mathcal{T}_m$ to predict 4-second motion tempo-density given the input 4-second music Mel-features. The Music Tempo-Density Estimator is composed of 2 transformer layers, 5 convolution blocks, and 5 deconvolution blocks. The training of $\mathcal{T}_m$ is based on a pre-trained $\mathcal{T}_d$ and 8-second segments $(X^t, m^t)$ extracted from paired Music-Dance Dataset $\mathcal{X}$. Only the tempo-density part of the $\mathcal{T}_d$ will be used, $X_{td} = \mathcal{T}_d(X)$. As $\mathcal{T}_m$ is designed to process 4-second segments, the middle 4 seconds of the input music will be the input and the middle 4 seconds of the tempo-density will be the target. By defining $\text{Middle()}$ as the function that extracts the middle 4 seconds from the 8-second sequence, the training loss is described below:
\begin{equation}
    \mathcal{L}_{tempo}(X^t, m^t) =  ||\mathcal{T}_m(\text{Middle}(m^t)) - \text{Middle}(\mathcal{T}_d(X^t))||_2
\end{equation}

The idea of predicting motion tempo-density based on music features is to make sure that the rhythmic features from music and motion are in the same modality and can be efficiently comparable. It is expected that after training, $\mathcal{T}_m$ will efficiently learn the temporal-relevant features from the music, which means the predicted motion tempo-density will show the expected tempo alignment for the music. The time-warped motion having a similar tempo-density is expected to be well-aligned to the music.

During inference, the same 4-second music segment set $\{m^t\}$ produced in Sec. \ref{sec:selector} will be processed by Music Tempo-Density Estimator $\mathcal{T}_m$ to predict the set of 4-second tempo-densities $\{\mathcal{T}_m(m^t)\}$. On the other hand, the selected motion segment set $\{X^t_k\}_{k=1}^{K}$ for each music segment $m^t$ will be processed by Motion Tempo-Density Estimator $\mathcal{T}_d$ to obtain 8-second tempo-density, while the motion content vector part in $\mathcal{T}_d$ will again be ignored.
Instead of using the Random Time-Warper $\mathcal{D}$, we now use Dynamic Time-Warper $\mathcal{W}$ to align predicted tempo-density from $\mathcal{T}_m$ and motion tempo-density from $\mathcal{T}_d$. $\mathcal{W}$ is based on subsequence extraction version of DTW algorithm \cite{sakoe1978dynamic,mcfee2015librosa} with slope constraint in $[0.5,2]$. Given the target 4-second tempo-density predicted from audio features $\mathcal{T}_m(m^t)$ and the 8-second source motion tempo-density from $\mathcal{T}_d(X^t_k)$ for each motion in the set, $\mathcal{W}$ produce an alignment path $P$, which extract a 4-second warped subsequence $\lceil X\rfloor^t_k$ from the 8-second motion sequence $X^t_k$. After warping all sequences in the selected motion segment set, we obtain a warped motion segment set $\{\lceil X\rfloor^t_k\}_{k=0}^{K-1}$.

\subsection{Dynamic graph for motion node transition}\label{sec:dynamicgraph}

The Dynamic Graph will organize motion segments into motion nodes. Similar to the Motion Graph, our Dynamic Graph motion node represents 1 second of motion, and our edge connecting the motion nodes represents the cost of node transition based on music style cost, action completeness cost, and transition cost.

We denote Dynamic Graph as $\mathcal{G}$ which contains $N_g$ numbers of 4-second motion segments selected at different times. $\mathcal{G}$ is maintained dynamically and updated every second.
We use $x_t^g$ to denote the motion node with index $g\in [1,N_g]$ at current time $t$, and we use $\{x^g\}_4$ to indicate the 4-second motion clip specified by the index $g$.

First, each warped motion segment in $\{\lceil X\rfloor_t^k\}_{k=0}^{K-1}$ from Sec. \ref{sec:embedding} based on music segment $m^t$ are organized into into 4 consecutive motion nodes of 1 second each, $\lceil X\rfloor= \{x_j\}_{j=1}^4$. As the inserted nodes of different $j$ are the candidate for music from time $t$ to $t+3$, we will activate the corresponding node when we select motion nodes for the corresponding music segment $\{m^t, ..., m^{t+3}\}$.
When we synthesize motion using the Dynamic Graph based on input music segments $\{m^t\}$, we select one motion node every second from the activated nodes based on the current music style and current motion node. The motion nodes that should be activated for $m^t$ are one of the nodes in each warped motion segment from time $t-3$ to $t$, so $N_g = K*4$ nodes will be considered every time, and motion segments inserted before $t-3$ will be removed from the graph.

Given that our current motion node is $\mathbf{x}_{t-1} = x^{g_0}_{t-1}$, an active motion node $x^{g_1}_{t}$, and current music style $s_m^t$, we calculate the cost of node transition between $x^{g_0}_{t-1}$ and $x^{g_1}_{t}$ as follows:

\begin{equation}
\begin{split}
    \mathcal{L}_{node}(x^{g_0}_{t-1}, x^{g_1}_{t}, s_m^t)& =  \lambda_s \mathcal{L}_{S}(x^{g_1}_{t}, s_m^t) + \lambda_c \mathcal{L}_{C}(x^{g_0}_{t-1}, x^{g_1}_{t}) \\
    & + \lambda_t \mathcal{L}_{T}(x^{g_0}_{t-1}, x^{g_1}_{t})
\end{split}
\end{equation}

\noindent\textbf{Music Style cost} $\mathcal{L}_{S}$ calculates the difference between the current music style embedding and the motion style embedding of the motion node calculated in Sec. \ref{sec:selector}. We assume the style evaluated for the whole motion segment is temporally consistent, so all motion nodes in the segment share the same motion style embedding. Due to the style-based selection in Sec \ref{sec:selector}, the motion nodes selected at the same time will have very similar style embedding, but the style queried in adjacent timestamps may have significant difference due to immense style change. Therefore, we reuse style embedding to consider the style difference between motion nodes queried in different timestamps. For motion node $x^{g_1}_{t} \in \{x^{g_1}\}_4$,

\begin{equation}
    \mathcal{L}_{S}(x^{g_1}_{t}, s_m^t) = ||\mathcal{S}_d(\{x^{g_1}\}_4) - s_m^t||_2
\end{equation}

\noindent\textbf{Action Completeness cost} $\mathcal{L}_{C}$ aims to check the action completeness in the current motion node to avoid unpleasant artifacts such as a sudden stop in spinning. We assume that no motion beats will be detected when the sequence is not complete. Therefore, we calculate the frame distance between the latest beat in the current node $x^{g_0}_{t-1}$ and the earliest beat in the active node $x^{g_1}_{t}$ if they come from the same warped sequence. If the distance between that 2 beats is large, the action performed is more likely to be not complete. In such cases, $\mathcal{L}_{C}$ will be low. Here, $\mathbf{V}(\cdot)$ defines the distance between the latest beat position of the first half and the first beat position in the second half of the concatenated 2-second sequence. Note that the distance should be offset by 20 (1 second) to standardize the distance.
\begin{equation}
\begin{split}
\mathcal{L}_{C}(x^{g_0}_{t-1}, x^{g_1}_{t})= \begin{cases}
    1 & \text{if $g_0 \neq g_1$}\\
    \text{Sigmoid}( 20-\mathbf{V}(x^{g_0}_{t-1}, x^{g_1}_{t})/5 ) & otherwise
    \end{cases}
\end{split}
\end{equation}

\begin{table*}[t]
\centering
\caption{Objective metrics for evaluating the realism and beat alignment of the generated motion. The best and second best performances are highlighted in red and blue, respectively.}
\begin{tabular}{|>{\centering\arraybackslash}m{2.4cm}|>{\centering\arraybackslash}m{2.4cm}|>{\centering\arraybackslash}m{2.2cm}|>{\centering\arraybackslash}m{2.2cm}|>{\centering\arraybackslash}m{2.2cm}|>{\centering\arraybackslash}m{2.2cm}|}
\hline
\multicolumn{2}{|c|}{} &AI Choreo & Transflower & Motion Graph & Ours\\\hline
\multirow{2}{*}{Realism Metrics} &FPD$\downarrow$ &9.8821 & \textcolor{blue}{9.8729} & 9.8762 &\textcolor{red}{9.8682}\\\cline{2-6}
&FMD$\downarrow$ &16.9582 & \textcolor{blue}{16.9538} & 16.9575 &\textcolor{red}{16.9452}\\\hline
Beat Alignment &motion2audio $\downarrow$ &121.2 &173.6 &\textcolor{blue}{120.7} &\textcolor{red}{87.8}\\\cline{2-6}
Metrics  &audio2motion $\downarrow$ &\textcolor{blue}{1051.6} &\textcolor{red}{281.9} & 5485.6 &1302.4\\\hline
\end{tabular}\label{tab:objective}
\end{table*}

\noindent\textbf{Motion Transition cost}  $\mathcal{L}_{T}$ calculates the joint coordinate difference between the last motion frame of current motion node and the first motion frame of the active motion node. When we select the first motion node, the transition cost is always 0.
\begin{equation}
\begin{split}
\mathcal{L}_{T}(x^{g_0}_{t-1}, x^{g_1}_{t})
    =\begin{cases}
    0 & \text{if $t=0$}\\
    || (x^{g_0}_{t-1})\rfloor - \lceil(x^{g_1}_{t}) ||_2 & otherwise
    \end{cases}
\end{split}
\end{equation}
Here $(\cdot)\rfloor$ and $\lceil(\cdot)$ are the operators to extract the last frame and the first frame of the motion sequence.

Finally, we select the motion node $\mathbf{x}_t$ from all the active nodes that have the minimum cost.
As the motion node selection is based on more than transition cost, the transition between nodes may not be smooth and optimal. We apply linear blending between each pair of motion nodes to generate predicted motion $\mathbf{\hat{X}}$.

\section{Experiment}
\vspace{1mm}
As the main focus of our model is to generate dance motion conditioned on music, we compare our model to the two state-of-the-art music-conditioned dance generation models, the Transflower \cite{valleperez2021transflower} and the AI Choreographer \cite{li2021learn}. We want to test the Dynamic Graph module's effectiveness, so we also compare it to a different setup of our model where the Dynamic Graph is replaced by Motion Graph \cite{kovar2008motiongraph}. Although the motion source sequences will still be selected by the Style-based Selector and warped by the Dynamic Time-Warper, only the first second of the warped 4-second segment will be inserted into the Motion Graph. Note that the main difference between the Motion Graph variant and the Dynamic Graph variant is the absence of previously warped nodes. As the subsequent nodes warped from previous timestamps are discarded in the Motion Graph, the completeness cost is inapplicable in the graph node selection. Also, as the nodes selected by the Style-based Selector in the current timestamp will not be compared to stored subsequent nodes that are selected and warped previously, the style difference between motion nodes is small, and thus we remove the style cost in the Motion Graph.
All models will be trained on AIST++ dataset \cite{li2021learn} train set and tested on the corresponding test set. The comparisons of our method with AI Choreographer and Transflower show the difference between handling style correlation, tempo alignment, and motion transition in a step-by-step pipeline approach and an end-to-end combined approach. The comparison of the Dynamic Graph and Motion Graph shows the influence of the action completeness cost and the stored subsequent motion nodes in the Dynamic Graph.

Apart from the comparison to other models, we also perform ablation studies to evaluate the Style Embedding module, the Re-Tempo module, and the Dynamic Graph in our ChoreoGraph framework. Evaluations and Visualizations show how each module contributes to improving the output motion quality. Finally, a user study is performed to compare the ChoreoGraph performance to other models from the human perspective on different aspects.

\subsection{Implementation Details}
We implement our framework in PyTorch \cite{paszke2019pytorch}. As described in Sec. \ref{sec:selector} and Sec. \ref{sec:embedding}, we first train the Style Embedding modules $\mathcal{S}_m$ and $\mathcal{S}_d$. Then, the Tempo-Density Estimators $\mathcal{T}_d$ and $\mathcal{T}_m$ are trained in two phases. First, we train motion Tempo-Density Estimator $\mathcal{T}_d$ for 10000 epochs. Next, we train music Tempo-Density Estimator $\mathcal{T}_m$ for 10000 epochs while keeping the weight of $\mathcal{T}_d$ fixed. Both training phases utilize ADAM \cite{kingma2014adam} optimizer with a batch size of 16 and a learning rate of $1 \times 10^{-4}$.

\subsection{Quantitative Evaluation}\label{sec:quanteval}
We quantitatively evaluate the quality of the generated dance in terms of motion realism and alignment with music beats.

\subsubsection{Motion Realism.}

We applied Fréchet pose distance (FPD) and the Fréchet movement distance (FMD) \cite{valleperez2021transflower} to measure the realism of generated motions among different methods. FPD and FMD measure the Fréchet distance of the distribution of poses $p_i$ and that of the distribution to the concatenations of three consecutive poses $(p_{i-1}, p_{i}, p_{i+1})$. Following the settings from \cite{valleperez2021transflower}, the measures are computed using the pose joint exponential map without normalization. For each test music piece, the distribution of generated motions of each method is compared to that of real data to check which model captures the distribution of real movements best. The results are shown in Table \ref{tab:objective}. It can be seen that our ChoreoGraph produces the smallest FPD and FMD of all competing methods.

\begin{table*}[t]
\centering
\caption{Ablation study measures the artist Label Distances between the music and the motion. Smaller Label Distances indicate closer aesthetic styles between music and motion. The better one is highlighted in red.}
\begin{tabular}{|>{\centering\arraybackslash}m{2.2cm}|>{\centering\arraybackslash}m{3cm}|>{\centering\arraybackslash}m{3cm}|>{\centering\arraybackslash}m{3cm}|>{\centering\arraybackslash}m{3cm}|}
\hline
&\textbf{Body openness} &\textbf{Motion intensity} &\textbf{Motion rhythm} &\textbf{Motion Asymmetry}\\
& vs.& vs.& vs.& vs.\\
&\textit{Intervallic structure $\downarrow$} &\textit{Rhythmic density $\downarrow$} &\textit{Time between onsets $\downarrow$} &\textit{Spectral contents $\downarrow$}\\\hline
\textit{Real Data } &1.9193  &2.0671 &1.2870 &1.1607\\\hline
\textit{Transflower} &1.9655  &1.9992 &1.2850 &\textcolor{blue}{1.1344}\\\hline
\textit{AI Choreo} &2.0501  &2.0541 &1.3583 &1.1663\\\hline
\textit{Motion Graph} &\textcolor{blue}{1.7768}  &\textcolor{blue}{1.9772} &\textcolor{blue}{1.2688} &1.2460\\\hline
\textit{Ours} &\textcolor{red}{1.7375}  &\textcolor{red}{1.8888} &\textcolor{red}{1.2380} &\textcolor{red}{1.0514}\\\hline
\end{tabular}\label{tab:styledist}
\end{table*}

\subsubsection{Music Beats Alignment.}\label{sec:beateval}

We also evaluate the music beat and motion beat alignment, which shows important rhythmic attributes. For the music beat extraction, we applied the beat check function in the Librosa \cite{mcfee2015librosa} music processing library. For the motion beat extraction, we extract the local minima of the motion speed. Similar to the AI Choreographer \cite{li2021learn} and Transflower \cite{valleperez2021transflower} approach, we calculate the Beat Alignment Score (BAS), which is the mean squared distance between every motion beat and its nearest music beat (motion2music), and that between every music beat and its nearest motion beat (music2motion). The results are shown in Table \ref{tab:objective}. It can be seen that our method significantly outperforms AI Choreo and Transflower in the motion2audio BAS, which is expected because we explicitly warp the motion based on learned tempo-density functions. The result of the Motion Graph also shows the effect of beat alignment from the Re-Tempo module as motion nodes are selected and warped every second. However, the processed motion nodes are unlikely to match with previously selected motion as they are warped differently, so the linear blending that smooths the motion node transition may smooth out some beat information and lower the beat alignment performance. In the audio2motion BAS score, however, Transflower performs better, since it generates motion based on every music window, while our ChoreoGraph imposes several constraints on the distortion that can be applied to the content motion. As the number of motion beats is usually less than the number of audio beats, there will be audio beats missed from time to time. However, Transflower tends to generate repeated motion just to align to the audio beats, which makes the dance less diverse and less natural, as to be validated in our user study.

\subsection{Ablation Study}

\begin{table}[t]
\centering
\caption{Motion-music feature pairs implemented for artistic evaluation. Note that each feature is evaluated on a scale of 1 to 10, and then associated with quantifiable metrics.}
\begin{tabular}{|p{0.44\linewidth} | p{0.47\linewidth}|}
\hline
\textbf{Dance Labels} & \textbf{Music Labels} \\\hline
\textbf{Body openness:} total distance between head, limbs to the body center
& \textbf{Intervallic structure:} distance between fundamental frequencies and pitch-sequence repetitiveness.\\\hline
\textbf{Motion intensity:} absolute mean of all joints' coordinate variations.
& \textbf{Rhythmic density:} number of onsets per second.\\\hline
\textbf{Motion rhythm and smoothness:} accumulated angular acceleration within a given time window.
& \textbf{Time between onsets} and rhythmic density.\\\hline
\textbf{Asymmetry:} independent use of only the upper or lower part of body.
& \textbf{Spectral contents:} the rough/ smooth and aggressive/ mild sound dualisms.\\\hline
\end{tabular}\label{tab:artlabel}
\end{table}
\subsubsection{The Style Embedding Module}\label{sec:ldeval}

As introduced in Sec. \ref{sec:selector}, we select the motion source sequences based on the trained style vector. As the style matching relation is implicit and hard to judge, we asked professional artists to design quantifiable artist labels, and we measured artist Label Distances (LD) between AIST++ dataset music and paired motion, as well as the generated motions from different methods. Table \ref{tab:artlabel} shows the feature labels which are incorporated as attributes to each data item in the MDR. Note that each of the labels can be associated with quantifiable metrics and then normalized to the same scale. The results are shown in Table \ref{tab:styledist}.

As can be seen, the LD becomes smaller for all 4 artist labels, especially on the \textit{motion intensity} vs. \textit{rhythmic density} and \textit{motion asymmetry} vs. \textit{spectral contents}. The reason for \textit{motion intensity} vs. \textit{rhythmic density} comes from our explicit Dynamic Time-Warper $\mathcal{W}$ on beat alignment. For the \textit{motion asymmetry} vs. \textit{spectral contents}, the selected input features for the Style Embedding modules, which are the joint coordinates and Mel-spectrogram, can represent and measure music spectral features and motion left-right body asymmetry easily. Our Style Embedding modules capture such artistic aspects and apply them to the style-based selection and Dynamic Graph style cost evaluation. The other LDs also indicate that the motion matches better the music style in our proposed method.

Apart from our ChoreoGraph result, we also notice that the LD for the Motion Graph version of our framework performs better than the other 2 methods. The reason is that the motion nodes are selected from the Style-based Selector and warped by Dynamic Time-Warper before inserting to the Motion Graph. This proves the performance of the Style-based Selector and Dynamic Time-Warper in generating style-matched and beat-aligned motion nodes.

\begin{table}[t]
\centering
\caption{Ablation study for Re-Tempo module on Beat Alignment Scores. For without Re-Tempo, the middle 4-second in the 8-second clip is selected instead of the warped segment}
\begin{tabular}{|>{\centering\arraybackslash}m{2.4cm}|>{\centering\arraybackslash}m{2.4cm}|>{\centering\arraybackslash}m{2.4cm}|}
\hline
BAS & \textit{without} Re-Tempo & \textit{with} Re-Tempo\\\hline
motion2audio $\downarrow$ &128.9 &\textcolor{red}{87.8}\\\hline
audio2motion $\downarrow$ &\textcolor{red}{735.1} &1302.4\\\hline
\end{tabular}\label{tab:ablation_retempo}
\end{table}

\begin{figure*}[t]
\setlength{\abovecaptionskip}{5mm}
\setlength{\belowcaptionskip}{5mm}
\centering
\includegraphics[width = 1\linewidth]{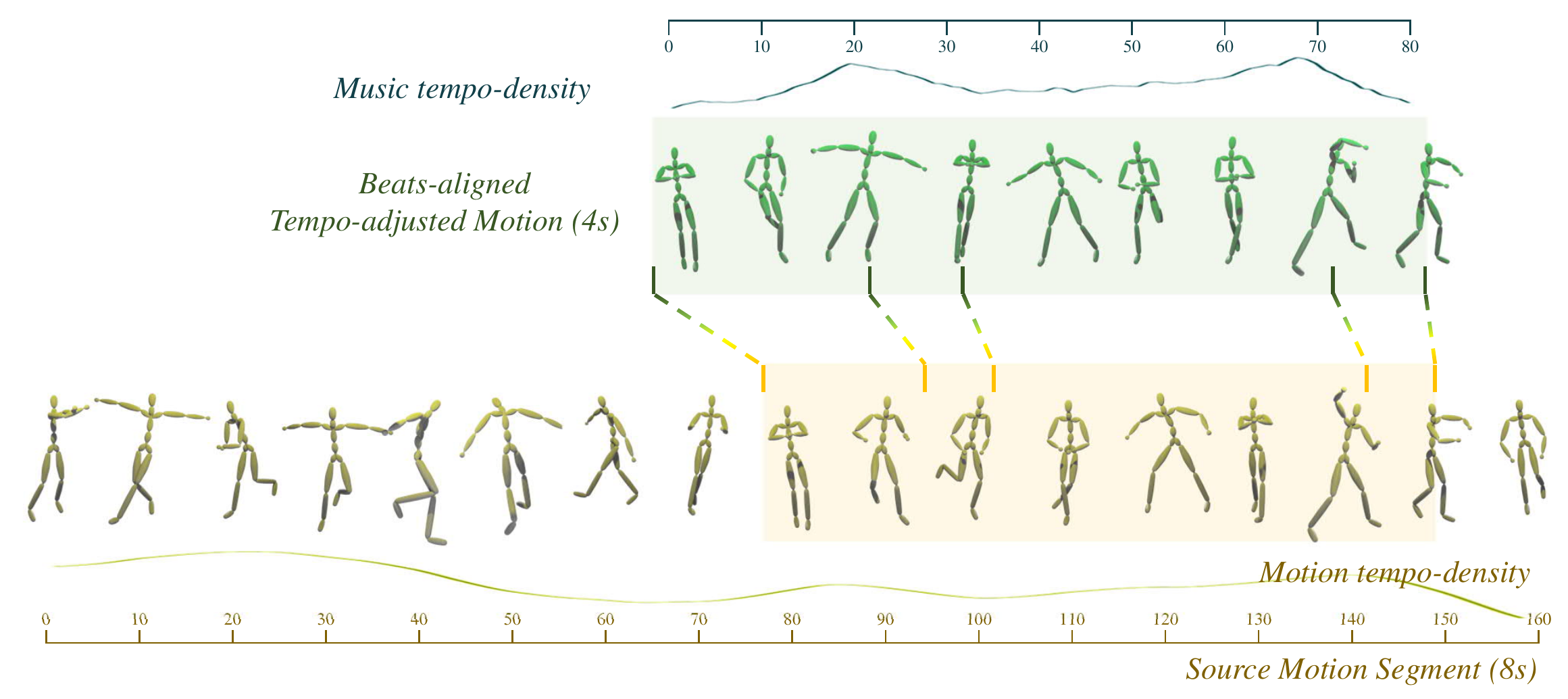}
\caption{Visual demonstration of the tempo adjustment and beats alignment function in the Dynamic Time-Warper $\mathcal{W}$. Note that the selection of motion frames is based on the tempo-density alignment.}
\label{fig:retempo}
\end{figure*}

\subsubsection{The Re-Tempo Module.}
To show the influence of the Re-Tempo module, we experiment and evaluate our model without the Dynamic Time-Warper $\mathcal{W}$ for comparison. For the framework without $\mathcal{W}$, as we evaluate the style of the 8-second motion source segment based on the middle 4 seconds of motion as described in Sec. \ref{sec:selector}, we extract the middle 4 seconds directly and process that into motion nodes. Then, we compare the beats alignment scores (BAS) between the music and the motion with and without the Re-Tempo module, the average BAS scores for the testing dance clips are shown in Table \ref{tab:ablation_retempo}. As we can see, the BAS scores for motion2audio have significantly improved as the Re-Tempo module performs explicit beat and tempo alignment to match each motion beat. As described in Sec. \ref{sec:beateval}, there will be audio beats without matched motion beats, so the audio2motion score becomes worse.

Next, we visually demonstrate the Beats alignment and tempo adjustment functions of the Re-Tempo module. The input 8-second source motion and output warped 4-second motion processed by Dynamic Time-Warper $\mathcal{W}$ are shown in yellow and green respectively in Fig. \ref{fig:retempo}. As can be seen, motion subsequence has been extracted from the source sequence and warped based on the difference between the Tempo-Density functions extracted from the motion and the music. Both the beats and the overall tempo have been modified to align with the music features. The effect of slope constraint in $\mathcal{W}$ can also be seen in the first 20 frames of the queried result, preventing the module from extracting too many frames and producing an abnormally fast motion sequence.

\subsubsection{The Dynamic Graph Module}
As described above, we replaced Dynamic Graph by Motion Graph for comparison, and we obtained better evaluation result using the Dynamic Graph. The beat alignment motion2audio in the Motion Graph is very good for almost all short sequences in the test set, but the performance turns bad as the music duration increases. This is caused by the increased percentage of linear interpolated frames for smoothing motion node transition, as the motion beats will be wiped out in the smoothing operation. On the other hand, our Dynamic Graph stores beat-aligned subsequent nodes that are warped together in previous timestamp with current motion node, and thus guaranteeing smooth transition and keeping the beat-alignment performance consistent with different input music duration.

\subsection{Human Evaluation}
Although we used different quantitative metrics to show our model performance, the evaluations on motion quality and its correlation to music are still not sufficient as the numeric metrics still cannot represent the generated motion quality clearly and effectively. Therefore, we evaluate our model performance with a user study. We prepared 10 music segments of 6 seconds each, and generate motions for each music segment using our proposed ChoreoGraph, and compare with those generated by the AI Choreographer \cite{li2021learn}, the Transflower \cite{valleperez2021transflower}, and Motion Graph version of our framework.

In the user study, each user watches 10 6-second videos generated by each method, and is asked to rate them on a scale between 0 and 5 (5 being the best quality) in four aspects described below:
\begin{itemize}
\item \textbf{Motion Variation.} Are the generated motion based on different music segments diverse without repeating.
\item \textbf{Motion Naturalness.} How natural and realistic is the dance motion. Is the dance motion generated by a machine or performed by a real human dancer.
\item \textbf{Style Match.}
Does the generated motion style match the music style (beat alignment should not be considered).
\item \textbf{Rhythm Alignment.}
Do the generated motion rhythm match the music beats.
\end{itemize}

\begin{figure}[t]
\centering
\includegraphics[width = 1\linewidth]{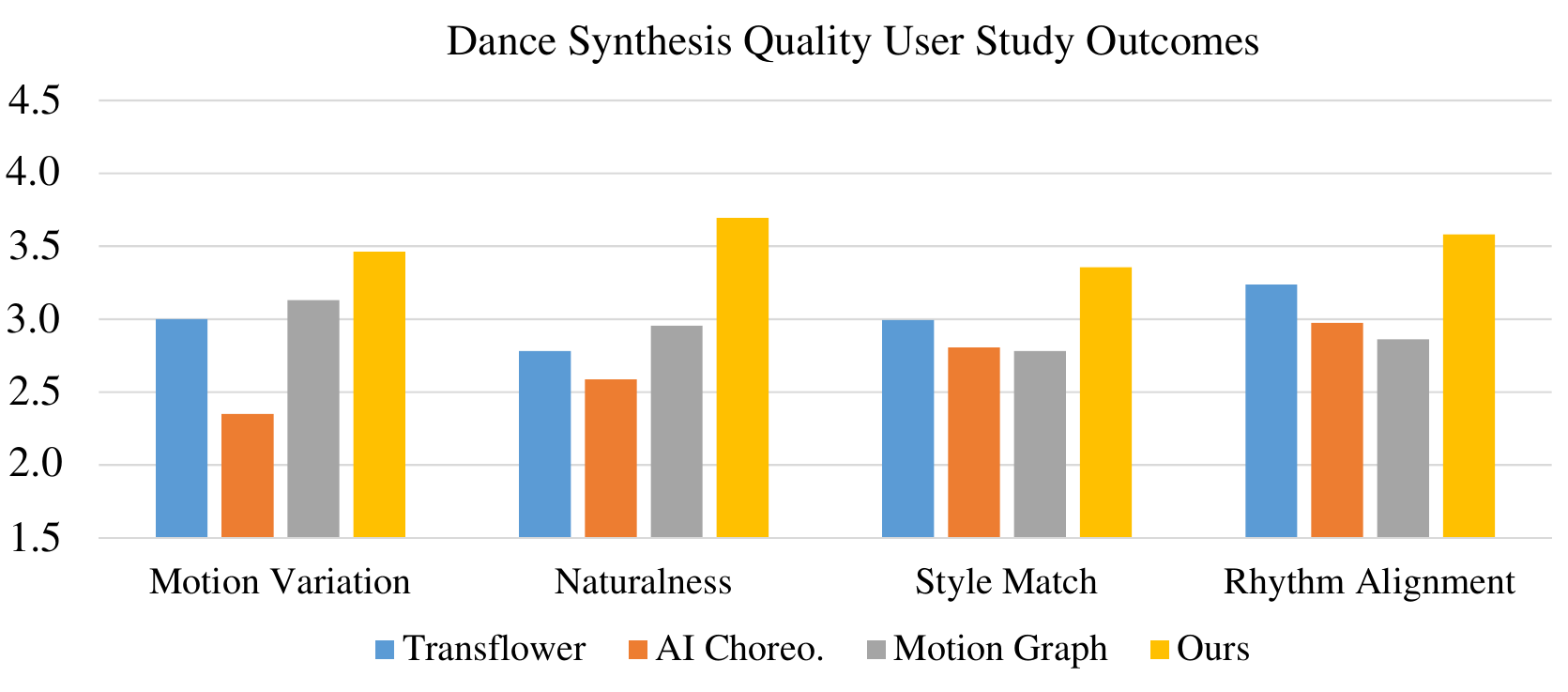}
\caption{Results of user evaluation on the generated motion.}
\label{fig:userstudy}
\end{figure}

We have collected answers from 20 independent judges, and the results are shown in Fig. \ref{fig:userstudy}. As can be seen, our proposed ChoreoGraph framework outperforms others significantly in all four aspects. The largest advantage is in motion naturalness which is expected since the generated motions from our framework are based on real motion sequences stored in the dataset $\mathcal{X}$. The node selection handling the beat alignment and action completeness also reduces the awkward dissonance between music and motion and makes the generated motion more natural and compatible with the input music.

\section{Conclusions}
\vspace{1mm}
We have proposed a dance generation framework, named ChoreoGraph, which selects and warps dance clips to synthesize high-quality dance motion for a given piece of music. A data-driven learning strategy is proposed to efficiently correlate the style and rhythmic connections between music and motion to enable the generation of beats-aligned motion nodes, which will be subsequently used in the Dynamic Graph to generate motion sequences with style and rhythm matched to input music. Quantitative results, qualitative results, and human evaluations demonstrated the efficiency of our proposed model. The proposed graph-based paradigm can generate motions with impressive motion quality and diversity.

\section*{Acknowledgments}
\vspace{1mm}
The research was supported by the Theme-based Research Scheme, Research Grants Council of Hong Kong (T45-205/21-N).

\vspace{5mm}

\bibliographystyle{ACM-Reference-Format}
\balance
\bibliography{2022.04_ACM_DanceGen}

\end{document}